\documentclass[iop, apj, twocolappendix, numberedappendix, appendixfloats]{emulateapj}
\usepackage{threeparttable}
\usepackage{multirow}
\usepackage{times}
\usepackage{enumitem}
\usepackage{url} 
\usepackage{amsmath,bm}
\usepackage{lineno}

\shorttitle{Parallax Zero-Point of \textit{Gaia} Early Data Release 3}
\shortauthors{Y. Huang et al.}

\begin{document}

\title{The Parallax Zero-Point of \textit{Gaia} Early Data Release 3 from LAMOST Primary Red Clump Stars}
\author{Yang Huang\altaffilmark{1}}
\author{Haibo Yuan\altaffilmark{2}}
\author{Timothy C. Beers\altaffilmark{3}}
\author{Huawei Zhang\altaffilmark{4, 5}}
\altaffiltext{1}{South-Western Institute for Astronomy Research, Yunnan University, Kunming 650500, People's Republic of China; {\it yanghuang@ynu.edu.cn}}
\altaffiltext{2}{Department of Astronomy, Beijing Normal University, Beijing 100875, People's Republic of China}
\altaffiltext{3}{Department of Physics and JINA Center for the Evolution of the Elements (JINA-CEE), University of Notre Dame, Notre Dame, IN 46556, USA; {\it tbeers@nd.edu}}
\altaffiltext{4}{Department of Astronomy, School of Physics, Peking University, Beijing 100871, P.\,R.\,China}
\altaffiltext{5}{Kavli Institute for Astronomy and Astrophysics, Peking University, Beijing 100871, P.\,R.\,China}

\begin{abstract}
We present an independent examination of the parallax zero-point of the Third {\it Gaia} Early Data Release (hereafter EDR3), using the LAMOST primary red clump (PRC) stellar sample.
A median parallax offset of around $26 \mu$as, slightly larger than that found by examination of distant quasars, is found for both the five- and six-parameter solutions in EDR3, based on samples of over {63,000} and 2000 PRC stars, respectively.
Similar to the previous investigation of Lindegren et al., to which we compare our results, the parallax zero-point exhibits clear dependencies on the $G$  magnitudes, colors, and positions of the objects.  
Based on our analysis, the zero-point of the revised parallax can be reduced to a few $\mu$as, and some significant patterns, e.g., discontinuities with stellar magnitude, can be properly removed.  However, relatively large offsets ($> 10 \mu$as) are still found for the revised parallaxes over different positions on the sky.
\end{abstract}

\section{Introduction}
\begin{figure*}
\begin{center}
\includegraphics[scale=0.425,angle=0]{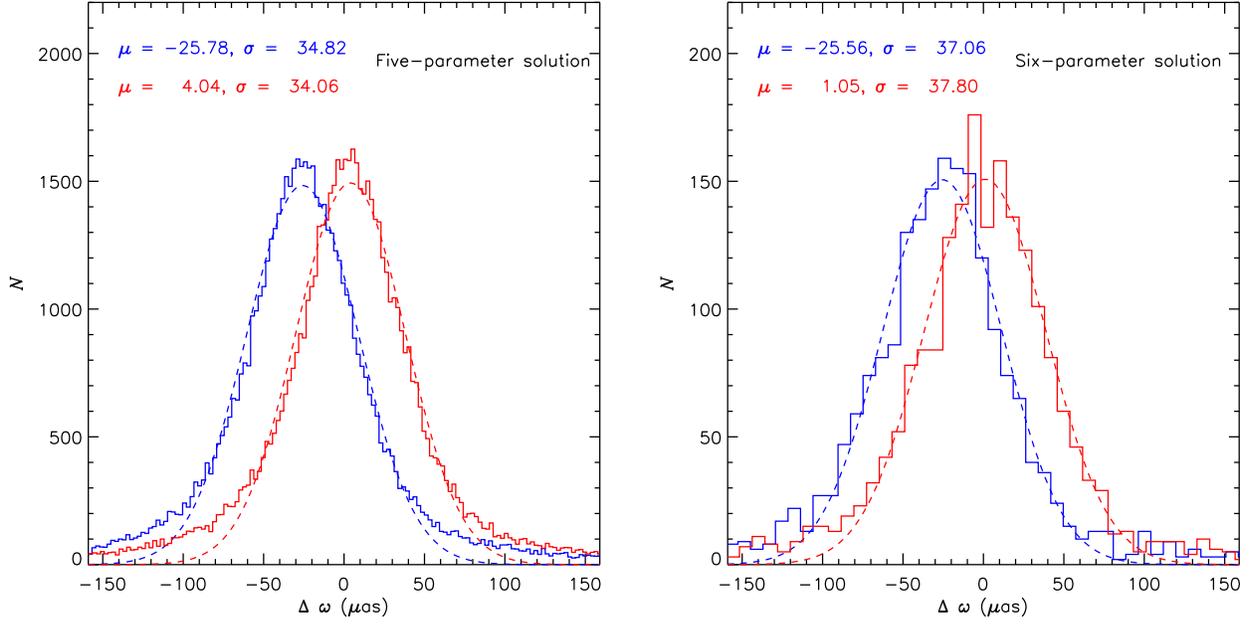}
\caption{{{\it Left panel:} Distribution of parallax differences between the \textit{Gaia} EDR3 five-parameter solution and the PRC sample. The blue histogram represents the difference distribution of $\omega_{\rm EDR3} - \omega_{\rm RC}$ and the red histogram represents the difference distribution of $\omega_{\rm EDR3}^{\rm Corr} - \omega_{\rm RC}$. $\omega_{\rm EDR3}^{\rm Corr}$ is the {\it Gaia} parallax with zero-point corrected using the model of Lindegren et al. (2020b). The blue and red dashed lines represent Gaussian fits for the two parallax difference distributions, respectively. The mean and standard deviation of the two distributions are labeled in the top-left corner with corresponding colors.
 {\it Right panel:} Similar to the left panel but for the {\it Gaia} EDR3 parallax from the six-parameter solution.}}
\end{center}
\end{figure*}

\begin{figure*}
\begin{center}
\includegraphics[scale=0.33,angle=0]{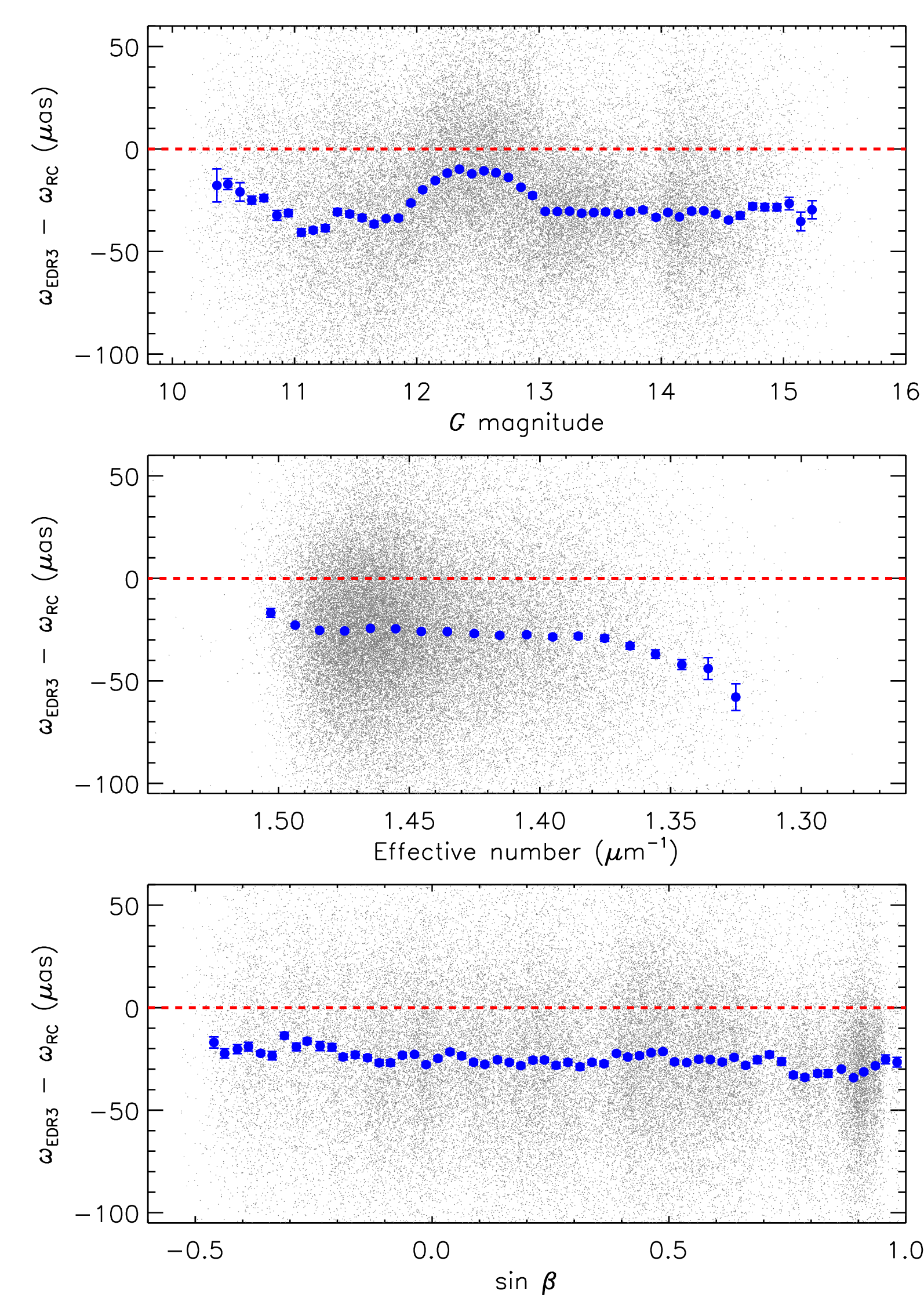}
\includegraphics[scale=0.33,angle=0]{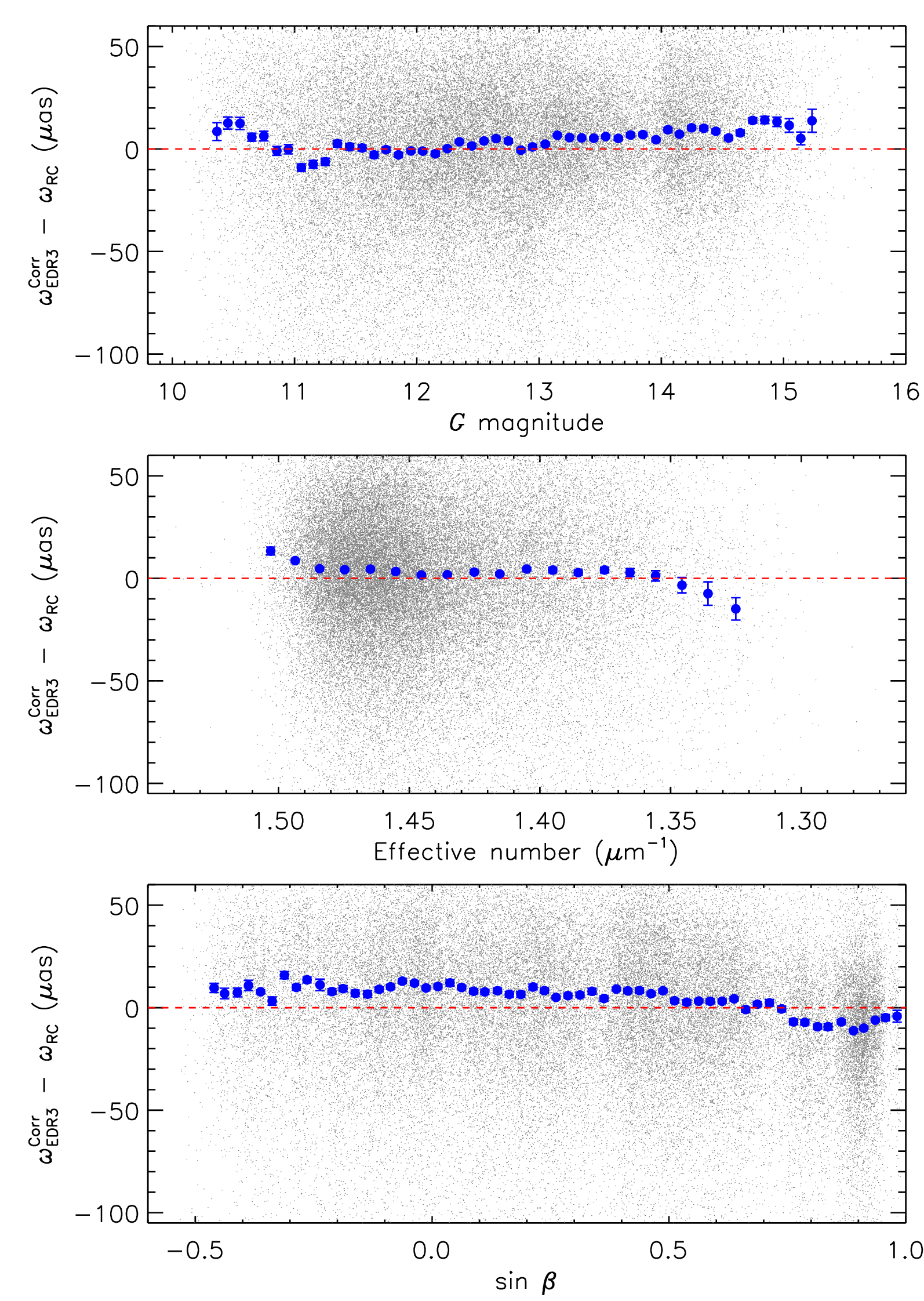}
\caption{{\it Left panels:} Parallax differences between the {\it Gaia} EDR3 five-parameter solution and the PRC sample, as a function of $G$ (top), effective wavenumber $\nu_{\rm eff}$ (middle), and ecliptic latitude (bottom).
The blue dots in each panel represent the median of the differences in the individual $G$, $\nu_{\rm eff}$, or ecliptic latitude bins. 
{The error bars represent 1$\sigma$ uncertainties of the median values (estimated with a bootstrapping procedure).}
The number of stars in each bin is no less than 100.
The red dashed lines in each panel mark the zero differences.
{\it Right panels:} Same as the left panels, but for the {\it Gaia} parallax with zero-point corrected using the model of Lindegren et al. (2020b).}
\end{center}
\end{figure*}

\begin{figure*}
\begin{center}
\includegraphics[scale=0.33,angle=0]{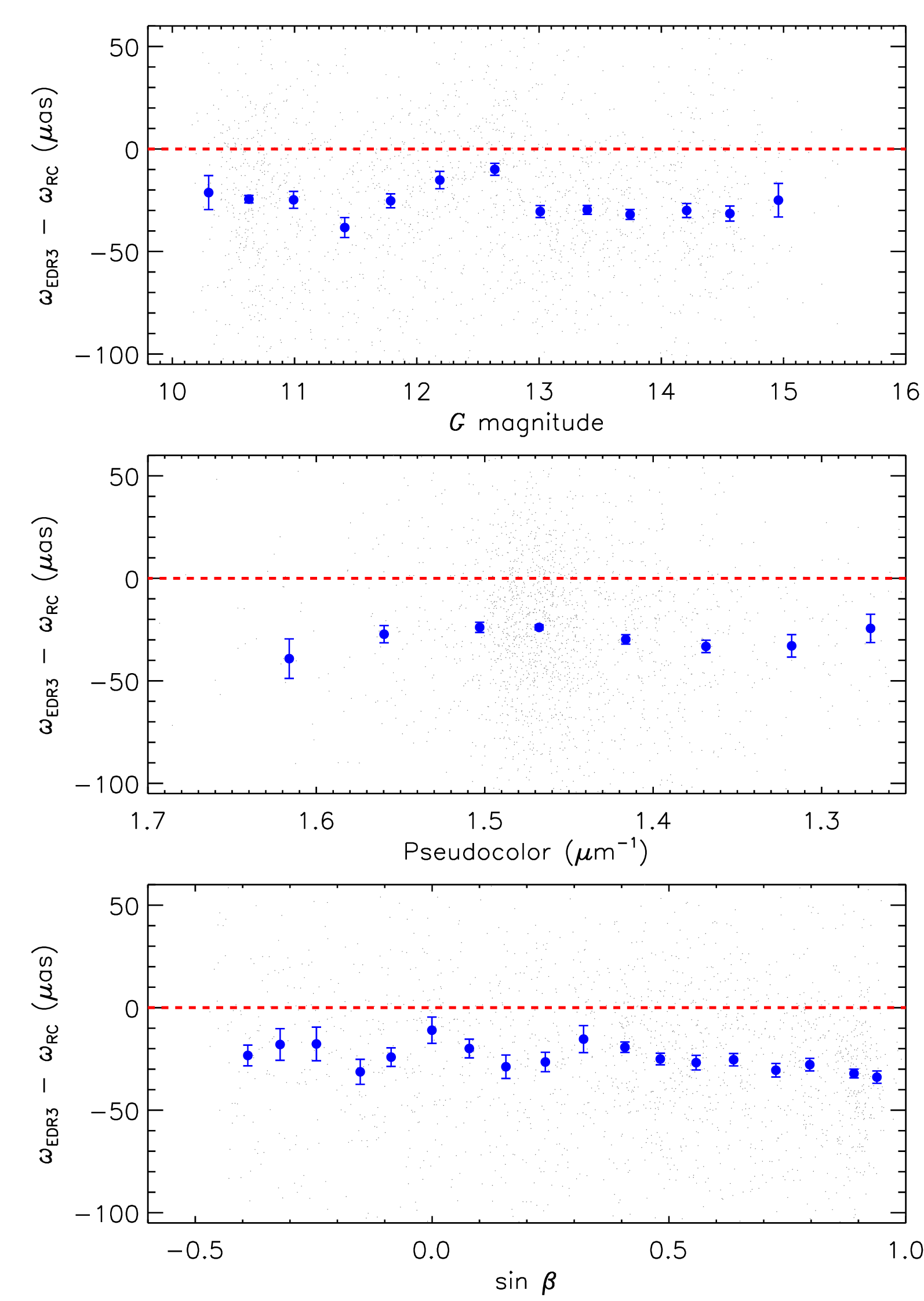}
\includegraphics[scale=0.33,angle=0]{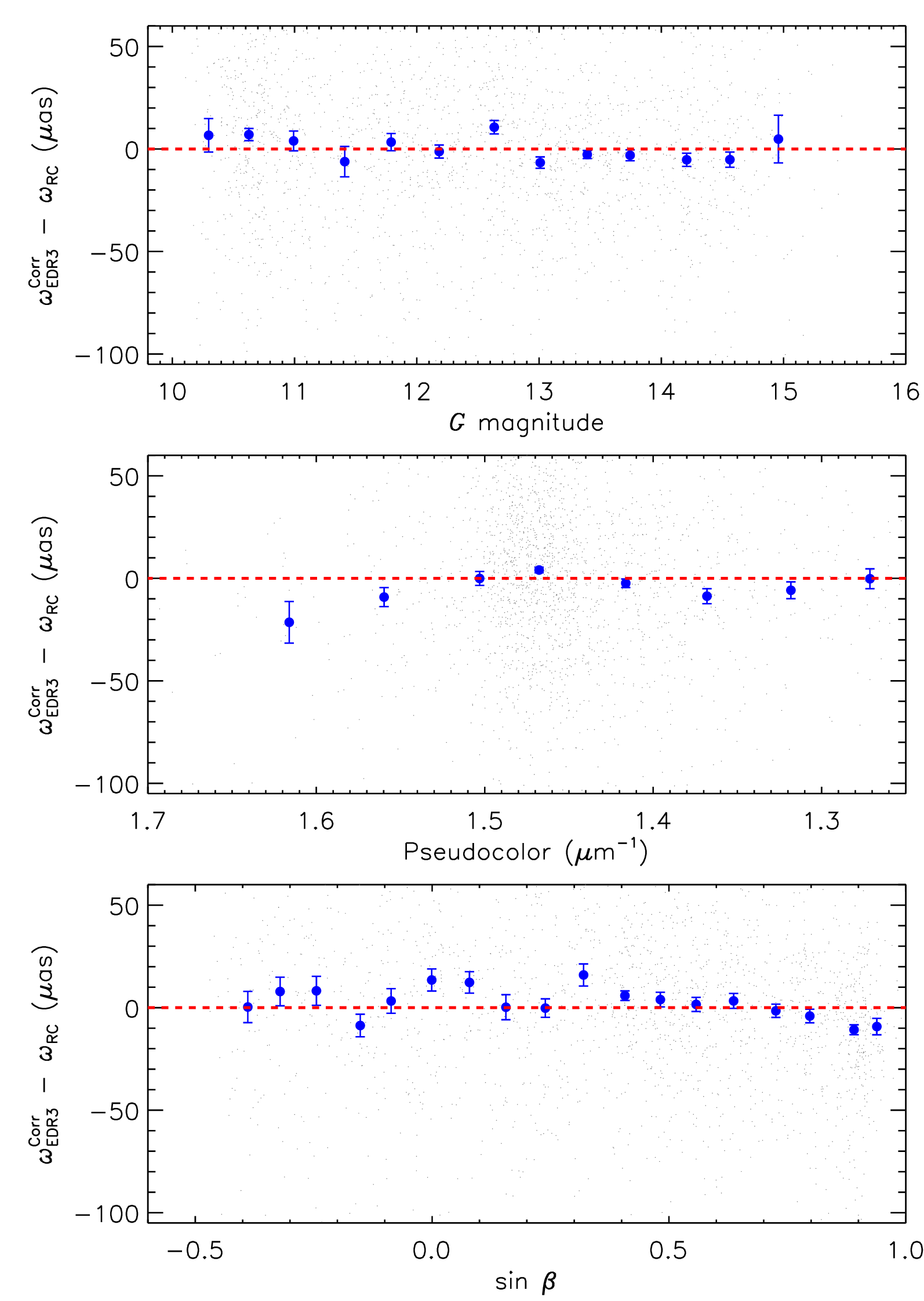}
\caption{Same as Fig.\,3, but for the {\it Gaia} EDR3 parallax from the six-parameter solution. Note that the central panels here show the parallax difference as a function of pseudocolor, rather than effective wavenumber (see text).  The number of stars in each bin is no less than 20.}
\end{center}
\end{figure*}

\begin{figure*}
\begin{center}
\includegraphics[scale=0.36,angle=0]{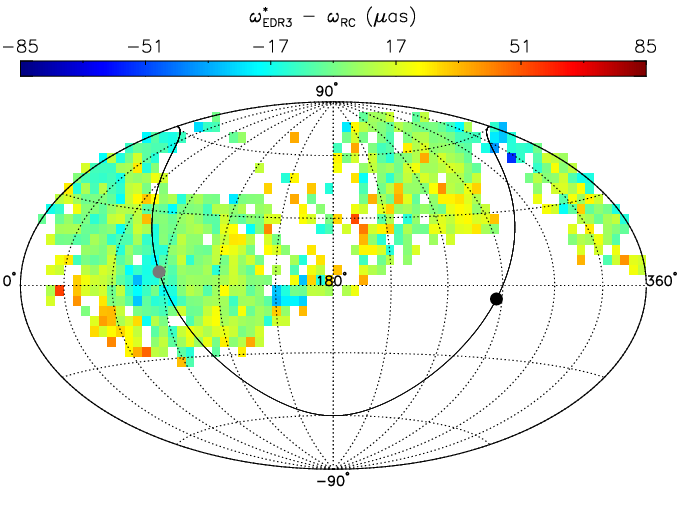}
\includegraphics[scale=0.36,angle=0]{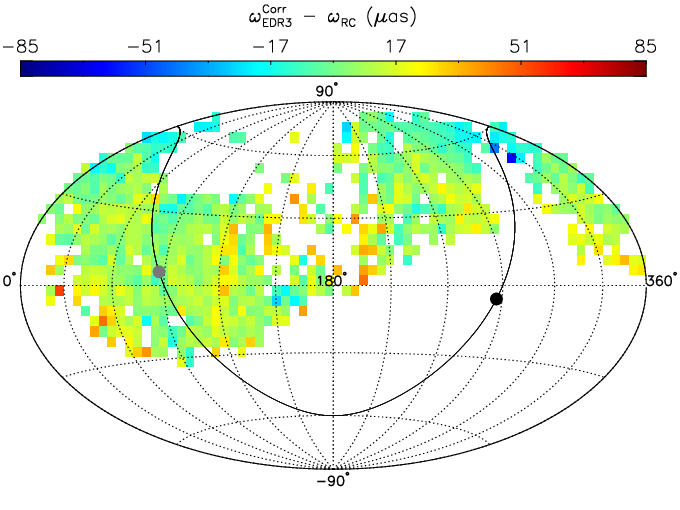}
\caption{Map in ecliptic coordinates of the median parallax difference between  $\omega_{\rm EDR3}^{*}$ and $\omega_{\rm RC}$ (left panel), and between $\omega_{\rm EDR3}^{\rm Corr}$ and $\omega_{\rm RC}$ (right panel), color-coded as indicated by the top color bars. 
To clearly show the position-dependent parallax bias, here $\omega_{\rm EDR3}^{*}$ represents the EDR3 parallax after correction for the magnitude dependence detected in Fig.\,3 (top-left panel).
$\omega_{\rm EDR3}$ here is given by the five-parameter solution. Each pixel covers an area of about $5 \time 5$ square degrees.
The number of stars in each pixel is no less than 15.
The solid lines in both panels mark the Galactic plane.
The black and grey dots mark the Galactic center and anti-center, respectively.}
\end{center}
\end{figure*}

\begin{figure}
\begin{center}
\includegraphics[scale=0.35,angle=0]{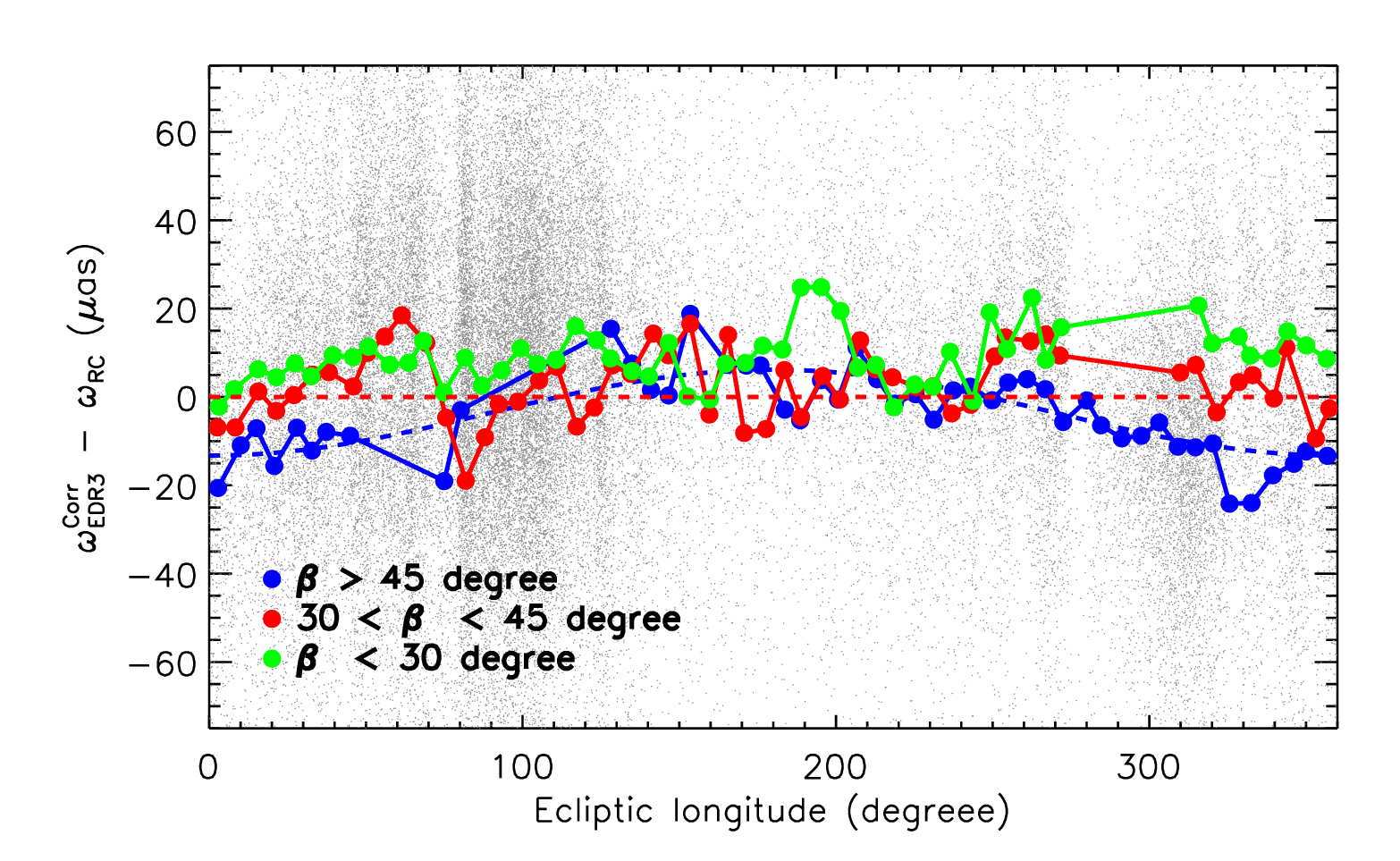}
\caption{Parallax difference between $\omega_{\rm EDR3}^{\rm Corr}$ and $\omega_{\rm RC}$, as a function of ecliptic longitude $\lambda$.
The blue, red, and green dots represent the median parallax difference in the individual $\lambda$ bins for ecliptic latitude $\beta > 45^{\circ}$, $30^{\circ} < \beta < 45$, and $\beta < 30^{\circ}$, respectively.
The number of stars in each bin is no less than 20.
{The blue dashed line represents a one-term Fourier model fit for the blue data points, $\Delta \omega = -3.5632 -9.7170\cos{(2\pi\phi)} + 0.0096\sin{(2\pi\phi)}$, where $\phi = \lambda/360^{\circ}$.}}
\end{center}
\end{figure}

Based on observations from the first 34 months (from July 2014 to May 2017) from the European Space Agency's {\it Gaia} mission (Gaia Collaboration et al. 2016), the  {\it Gaia} Early Data Release 3 (EDR3; Gaia Collaboration et al. 2020a) just released astrometric and photometric data for over 1.8 billion sources.
Of importance here, five astrometric parameters (position, parallax, and proper motions) are measured for over 81\% of these sources (1.468 billion) in EDR3 (Lindegren et al. 2020a).
These precisions are improved by 30\% for parallaxes, and by a factor of two for proper motions, respectively, when compared to those in {\it Gaia} DR2 (Gaia Collaboration et al. 2018).

Investigation of the systematic bias of the measured parallax in EDR3 is very important for its further applications, especially for estimating distances to the more distant stars.
The {\it Gaia} team has constructed a parallax zero-point model, depending on the $G$-band magnitude, spectral shape (colors), and ecliptic latitude of the sources {(Lindegren et al. 2020b)}. 
This model is derived based on a comprehensive analysis of quasars, binary stars, and stellar sources in the Large Magellanic Cloud (LMC). Evaluation of the 
parallax-bias dependencies for the three independent variables are different for sources with five- and six-parameter solutions\footnote{The descriptions of the five- and six-parameter solutions are given by Lindegren et al. (2020b). Briefly, a five-parameter solution is determined if the target has an accurate value of effective wavenumber ($\nu_{\rm eff}$) to select proper point-spread function (PSF) or line-spread function (LSF) in the data processing pipeline. For six-parameter solution, PSF or LSF at a default $\nu_{\rm eff} = 1.43 \mu$m$^{-1}$ is adopted due to no accurate value of $\nu_{\rm eff}$ of the concerned source.}.
For faint sources, the parallax bias is estimated directly by a large number of quasars covering almost the entire sky (except for the Galactic disk region).
However, for brighter sources and also sources with wider color ranges, the parallax bias is derived by an indirect way, based on physical binary stars and stars in the LMC.

In this letter, we perform an independent check on the parallax bias of EDR3 by using over {65,000} primary red clump (PRC) stars (Huang et al. 2020), identified from the LAMOST Galactic surveys (Deng et al. 2012; Liu et al. 2014).  This provides an important test of the derived zero-points that need to be used for correction of the EDR3 stellar parallaxes and their application to problems of contemporary interest. 
This paper is structured as follows. We introduce the data employed in Section\,2, and the main results are presented and discussed in Section\,3.
Finally, Section\,4 presents a summary of our results. 

\section{Data}
The current work adopts the PRC sample built by Huang et al. (2020; hereafter H20). With the information provided by high-quality {\it Kepler} asteroseismology data, the PRC stars were selected based on their positions in the 
metallicity-dependent effective temperature-surface gravity and color-metallicity diagrams (See Fig.\,1 of H20).
In total, nearly 140,000 PRC stars with spectral signal-to-noise ratios (SNR) higher than 20 were identified from the LAMOST Galactic surveys.
In addition, stellar masses and ages are derived for those PRC stars using the kernel principal-component method.
{Unlike assuming a constant absolute magnitude for PRC stars, as in previous work, H20 performed a new calibration of the $K_{\rm s}$ absolute magnitude for PRC stars, for the first time, by considering both the metallicity and age dependencies, using over 10,000 PRC stars with accurate distance estimates from Sch{\"o}nrich, McMillan \& Eyer (2019; hereafter SME19). 
By application of the statistical method developed by Sch{\"o}nrich, Binney \& Asplund (2012; hereafter SBA12), SME19 find an average parallax offset of $-54 \pm 6 \mu$as for {\it Gaia} DR2.
They then present accurate Bayesian distance estimates (from the parallax measurements with offset properly corrected) for the full {\it Gaia} DR2 RVS sample stars (Katz et al. 2018).
The distances for $>$10,000 PRC stars are taken from the RVS sample further by applying quality cuts as suggested by SME19 (see their Section\,8). 
The distances derived in this manner are accurate to a few per cent, as confirmed by their statistical method (SBA12).}
{ Moreover,} the distances of the PRC stars from this new calibration are also validated by the statistical method of SBA12 in H20. {The test indicates that the distances (from the new calibration) are very accurately determined (to better than a few per cent) for all concerned populations with  distances larger than 0.85\,kpc} (see Figs.\,13 and A1 of H20).

The LAMOST PRC sample is then matched with {\it Gaia} EDR3, and the following cuts are made to control the data quality:
\begin{enumerate}[label=\arabic*)]

\item LAMOST spectral SNR\,$\ge 50$, {distance larger than 0.85\,kpc,} stellar age $2 \le \tau \le 14$\,Gyr, and 2MASS $K_{\rm s}$  uncertainty smaller than 0.03\,mag;

\item The Renormalised Unit Weight Error (RUWE)\,$\leq 1.4$;

\item An effective wavenumber  $1.1 \leq \nu_{\rm eff} \leq 1.9$ for the five-parameter solution or $1.24 \leq$\,pseudocolor\,$\leq 1.72$ for the six-parameter solution.

\end{enumerate}

{The RUWE is an statistical indicator of the quality of the {\it Gaia} data, given by the re-normalised square root of the reduced chi-square of the {\it Gaia} astrometric solution.}
The definitions of effective wavenumber and pseudocolor are described in Lindegren et al. (2020a). 
Generally, they represent the color information of the observed targets.
The above cuts yield {63,138 and 2074} PRC stars with five- and six-parameter solutions in {\it Gaia} EDR3, respectively.
If not specified otherwise,  $\omega_{\rm RC}$, $\omega_{\rm EDR3}$, and $\omega_{\rm EDR3}^{\rm Corr}$ represent the parallaxes given by the distances of PRC stars in H20 ($1/d_{\rm RC}$), the parallaxes in EDR3, and the parallaxes in EDR3 with zero-point corrections from the code provided by Lindegren et al. (2020b), respectively.

 \section{Results and Discussion}
 
 The comparison between  $\omega_{\rm EDR3}$ and $\omega_{\rm RC}$  for the five- and six-parameter solutions are shown in {Fig.\,1}.
 The median offsets are all around $-26 \mu$as, slightly larger than the value found from distant quasars (Lindegren et al. 2020b).
 The parallaxes of EDR3 with zero-point corrections using the model of Lindegren et al. (2020b) are also checked, and the results are shown in {Fig.\,1}.
The median values of the parallax difference between $\omega_{\rm EDR3}^{\rm Corr}$  and $\omega_{\rm RC}$ are only {4.0 and 1.1}\,$\mu$as for the five- and six-parameter solutions, respectively.
These results confirm that the official parallax zero-point correction model can generally remove the global bias of EDR3 parallaxes.

To further test the main dependencies of the parallax bias in EDR3, {Figs.\,2 and 3} show the parallax difference between  $\omega_{\rm EDR3}$ and $\omega_{\rm RC}$, as a function of $G$, effective wavenumber/pseudocolor, and $\sin \beta$ (where $\beta$ represents the ecliptic latitude) for the five- and six-parameter solutions, respectively.
Thanks to the large number of PRC stars in the five-parameter solution in EDR3, the parallax-bias dependencies could be mapped in detail.
As shown in {Fig.\,2}, significant patterns are found for the EDR3 parallaxes as a function of $G$ magnitude.
The discontinuous ``jump-like" features found for $10 < G < 11$ and $12 < G < 13$ ranges are similar to those reported in Fabricius et al. (2020) and Lindegren et al. (2020b).
This check also shows a positive trend for parallax bias with effective wavenumber, especially at the red end ($\nu_{\rm eff} < 1.40$\,$\mu$m), as also reported in Fabricius et al. (2020) and Lindegren et al. (2020b). 
For ecliptic latitude, no significant trend is found, except a clear dip feature around $50^{\circ} < \beta < 70^{\circ}$.
For parallaxes derived from the six-parameter solution, the number of PRC stars is very limited, and most of the fine structures as a function of $G$ magnitude, pseudocolor, and $\sin \beta$ cannot be mapped with high accuracy.
We only note that the jump-like features in the parallax bias with $G$ magnitude can still be (marginally) seen.

The dependencies of the corrected EDR3 parallax bias are also checked; the results shown in {Figs. 2 and 3} for the five- and six-parameter solutions, respectively.
From inspection of {Fig.\,2}, the parallax biases (from the five-parameter solution), as a function of $G$ and effective wavenumber, are considerably reduced by the official zero-point correction model (Lindegren et al. 2020b).
{Most recently, the independent work by Zinn et al. (2021) has found similar results by using 2000 first-ascent red giant branch stars with asteroseismic parallaxes in the {\it Kepler} field.
Here, we note that the offset of the parallax difference for $G < 10.8$ and $G > 14.0$ are $9.8 \pm 1.0 \mu$as and $9.0 \pm 0.4 \mu$as, respectively. This implies that the corrected {\it Gaia} parallax may over-correct the zero-point at $G < 10.8$ and $G > 14.0$.
The over-estimated value of the parallax zero-point by the official {\it Gaia} model for $G < 10.8$ found here is similar to the values of $14 \pm 6 \mu$as given by  Riess et al. (2020) and $15 \pm 3 \mu$as reported by Zinn et al. (2021).}
For ecliptic latitude, the dip feature around $50^{\circ} < \beta < 70^{\circ}$ is still present.
For the six-parameter solution shown in {Fig.\,3}, the pattern of the parallax zero-point discontinuities with $G$ magnitude is largely removed.

In the official parallax zero-point correction model (Lindegren et al. 2020b), the position-dependent bias of EDR3 parallax for bright stars is not well-explored.
Here, {Fig.\,4} shows the 2D position-dependent bias of EDR3 parallaxes (from the five-parameter solution) in ecliptic coordinates.
Significant variations ($> 10 \mu$as) over the sky region covered by our PRC sample are clearly seen.
To obtain an estimate of the systematic zero-point dependence on positions, the parallax difference between  $\omega_{\rm EDR3}$ and $\omega_{\rm RC}$, as a 
function of ecliptic longitude $\lambda$, is shown in {Fig.\,5}.
For stars with ecliptic latitude $\beta > 45^{\circ}$, the median differences for individual $\lambda$ bins exhibit a clear trend, {well-described by a one-term Fourier model function (listed in the figure caption).}
For stars with $30^{\circ} < \beta < 45^{\circ}$, the median differences as function of $\lambda$ oscillate around zero with no significant trend.
For stars  with $\beta < 30^{\circ}$,  the median differences as a function of $\lambda$ oscillate around a positive global offset of $8 \mu$as, again without any systematic trend.
Given the similar sky coverage between the LAMOST observations and our PRC sample, we recommend correcting for the position-dependent bias of the EDR3 parallax found here when deriving distances from EDR3 parallaxes for LAMOST stars, and other samples covering similar regions of sky.

\section{Summary}

In this letter, both the EDR3 parallaxes and the revised EDR parallaxes obtained by the official zero-point correction model are checked independently by a sample of LAMOST PRC stars, which are believed to be good standard candles with distance accuracy better than 5 per cent.
With over {65,000} PRC stars, the global median offset of EDR3 parallax is found to be around $-26 \mu$as for both the five- and six-parameter solutions.
The parallax bias in EDR3 exhibits significant systematic trends with $G$ magnitudes, spectral shape (color), and positions.
The global offset and the main bias dependencies can be largely reduced by application of the official zero-point correction model.
However, the remaining biases in positions still remain, with variation amplitudes larger than 10\,$\mu$as.

\section*{Acknowledgements} 
{We would like to thank the referee for his/her helpful comments.}
It is a pleasure to thank Prof. Xiaowei Liu for a thorough read of the manuscript and helpful comments.
{We gratefully thank Joel C. Zinn for helpful discussions.}
This work is supported by National Key R\&D Program of China No. 2019YFA0405500 and National Natural Science Foundation of China grants 11903027, 11833006, 11973001, 11603002, 11811530289 and U1731108. 
Y. H. is supported by the Yunnan University grant C176220100006. 
We used data from the European Space Agency mission Gaia (\url{http://www.cosmos.esa.int/gaia}), processed by the Gaia Data Processing and Analysis Consortium (DPAC; see \url{http://www.cosmos.esa.int/web/gaia/dpac/consortium}). 
T.C.B. acknowledges partial support from grant PHY 14-30152, Physics
Frontier Center/JINA Center for the Evolution of the
Elements (JINA-CEE), awarded by the US National Science
Foundation. His participation in this work was initiated by conversations that took place during a visit to China in 2019, supported by a PIFI Distinguished Scientist award from the Chinese Academy of Science.

The Guoshoujing Telescope (the Large Sky Area Multi-Object Fiber Spectroscopic Telescope, LAMOST) is a National
Major Scientific Project built by the Chinese Academy of Sciences. Funding for the project has been provided by the
National Development and Reform Commission. LAMOST is operated and managed by the National Astronomical Observatories, Chinese Academy of Sciences.


\begin{thebibliography}{}
\bibitem[Deng et al.(2012)]{2012RAA....12..735D} Deng, L.-C., Newberg, H.~J., Liu, C., et al.\ 2012, Research in Astronomy and Astrophysics, 12, 735 
\bibitem[Fabricius et al.(2020)]{2020arXiv201206242F} Fabricius, C., Luri, X., Arenou, F., et al.\ 2020, arXiv:2012.06242
\bibitem[Gaia Collaboration et al.(2016)]{2016A&A...595A...1G} Gaia Collaboration, Prusti, T., de Bruijne, J.~H.~J., et al.\ 2016, \aap, 595, A1
\bibitem[Gaia Collaboration et al.(2018)]{2018A&A...616A...1G} Gaia Collaboration, Brown, A.~G.~A., Vallenari, A., et al.\ 2018, \aap, 616, A1
\bibitem[Gaia Collaboration et al.(2020)]{2020arXiv201201533G} Gaia Collaboration, Brown, A.~G.~A., Vallenari, A., et al.\ 2020, arXiv:2012.01533
\bibitem[Huang et al.(2020)]{2020ApJS..249...29H} Huang, Y., Sch{\"o}nrich, R., Zhang, H., et al.\ 2020, \apjs, 249, 29
{\bibitem[Katz et al.(2019)]{2019A&A...622A.205K} Katz, D., Sartoretti, P., Cropper, M., et al.\ 2019, \aap, 622, A205}
{ \bibitem[Lindegren et al.(2018)]{2018A&A...616A...2L} Lindegren, L., Hern{\'a}ndez, J., Bombrun, A., et al.\ 2018, \aap, 616, A2}
\bibitem[Lindegren et al.(2020)]{2020arXiv201203380L}
Lindegren, L., Klioner, S.~A., Hern{\'a}ndez, J., et al.\ 2020a, arXiv:2012.03380
\bibitem[Lindegren et al.(2020)]{2020arXiv201201742L} Lindegren, L., Bastian, U., Biermann, M., et al.\ 2020b, arXiv:2012.01742
\bibitem[Liu et al.(2014)]{2014IAUS..298..310L} Liu X. -W., et al., 2014, in Feltzing S., Zhao G., Walton N., Whitelock P., eds, Proc. IAU Symp. 298, Setting the scene for Gaia and LAMOST, Cambridge University Press, pp. 310-321, preprint (arXiv: 1306.5376)
{\bibitem[Riess et al.(2020)]{2020arXiv201208534R} Riess, A.~G., Casertano, S., Yuan, W., et al.\ 2020, arXiv:2012.08534}
\bibitem[Sch{\"o}nrich et al.(2012)]{2012MNRAS.420.1281S} Sch{\"o}nrich, R., Binney, J., \& Asplund, M.\ 2012, \mnras, 420, 1281
\bibitem[Sch{\"o}nrich et al.(2019)]{2019MNRAS.487.3568S} Sch{\"o}nrich, R., McMillan, P., \& Eyer, L.\ 2019, \mnras, 487, 3568
{\bibitem[Zinn(2021)]{2021arXiv210107252Z} Zinn, J.~C.\ 2021, arXiv:2101.07252}

\end{thebibliography}
\end{document}